\documentclass[12pt]{article}
\usepackage{amsmath,amsfonts}
\usepackage{color}

\numberwithin{equation}{section}

\let\ACMmaketitle=\maketitle
\renewcommand{\maketitle}{\begingroup\let\footnote=\thanks \ACMmaketitle\endgroup}

\begin{document}

\title{Non-Abelian gauge theories invariant under diffeomorphisms}
\author{Olivera Miskovic $^{1,}$\thanks{olivera.miskovic@pucv.cl} \ and Tatjana Vuka\v{s}inac $^{2,3,}$\thanks{tatjana@umich.mx} \bigskip \\
{\small $^1$ Instituto de F\'\i sica, Pontificia Universidad Cat\'olica de Valpara\'\i so,}\\
{\small Casilla 4059, Valpara\'{\i}so, Chile}\\
{\small $^2$ Facultad de Ingenier\'ia Civil, Universidad Michoacana de San Nicol\'as de Hidalgo,}\\
{\small Morelia, Michoac\'an 58000, Mexico}\\
{\small $^3$ Centro de Ciencias Matem\'aticas, Universidad Nacional Aut\'onoma de
M\'exico,}\\
{\small UNAM-Campus Morelia, A. Postal 61-3, Morelia, Michoac\'an 58090,
Mexico} }
\maketitle

\begin{abstract}

We discuss diffeomorphism and gauge invariant theories in three dimensions motivated by the fact that some models of interest do not have a suitable action description yet. The  construction is based
on a canonical representation of symmetry generators and on building of the corresponding canonical action.
We obtain a class of theories whose number of local degrees of freedom  depends on the dimension of the gauge group and the number of the independent constraints. By choosing the latter, we focus on three special cases, starting with a theory with maximal local number of degrees of freedom and finishing with a theory with zero degrees of freedom (Chern-Simons).

\end{abstract}

\section{Introduction}

Building a suitable theory which would provide, through the principle of least action, a dynamical description of a physical system of interest, has been
a long-standing problem in theoretical physics. Once having the action, one
can use well-established procedures to address different questions such as the existence of
symmetries, conserved charges, count of degrees of freedom, interactions, quantization of the theory, its renormalization, etc.,
especially the ones relevant off-shell.
The problem becomes even more important if one considers effective actions,
which capture effective dynamics of the system restricted to particular observable quantities.
The least action principle remains central in modern physics and mathematics.

However, many physically interesting models still lack their actions, and
are studied at the level of equations of motion, or are only defined as quantum theories.
For example,
higher-spin theories, which are quantum  theories of massless fields with
spin greater or equal to two, do not have action principle formulated yet,
except for some particular toy models. One of them is a three-dimensional
higher-spin gravity \cite{Banados:2012ue}, whose fields of spin $n\geq 2$
transform as the adjoint representations of the gauge group $SL(n,\mathbb{R})\times SL(n,\mathbb{R})$.
Gravitational field is one of them because AdS$_{3}\simeq SL(2,\mathbb{R})\times SL(2,\mathbb{R})$
is a subgroup of the full gauge group. An important feature of this action is that it is invariant
under two local symmetries which are usually difficult to unite -- namely,
spacetime diffeomorphisms and non-Abelian gauge symmetry.

Other examples that have drawn recent attention of the high energy physics community are
motivated by the success of AdS/CFT correspondence in description of
strongly-coupled field theories, in particular condensed-matter systems \cite{Sachdev:2010ch}.
They are usually
non-relativistic and their fields may scale differently with respect to time and space reparametrizations,
thus their effective theories are dominated by non-relativistic symmetries
described, for example, by the Schr\"{o}dinger algebra \cite{Son:2008ye,Balasubramanian:2008dm}
in the case of asymmetrical time and space scaling, or the Galilean algebra \cite{Bagchi:2009my}
in the symmetric case. Newton-Cartan gravity \cite{Cartan:1923zea,Misner} is a geometric version of
Newtonian gravity, constructed through gauging of a centrally extended
form of the Galilean algebra (Bargmann algebra) \cite{Andringa:2010it}.
However, until recently it did not have a suitable developed action principle, when it was constructed with the help of auxiliary gauge fields in the form of the Extended String Newton-Cartan Gravity \cite{Bergshoeff:2018vfn}. This example shows a complexity of formulation of an action invariant under a given non-Abelian symmetry.

On the other hand, one of the nice predictions of the string theory is an existence of a discrete family of quantum field theories in six dimensions invariant under a non-Abelian group symmetry described by the largest superconformal algebra that does not contain higher spin particles, known as the (2,0) algebra. The relevance of these field theories is that a large class of lower-dimensional supersymmetric field theories can be obtained geometrically by their compactifications, such as $N = 2$ superconformal field theories in four \cite{Gaiotto:2009we} and three \cite{Dimofte:2011ju} dimensions. (For more general discussion about other possible theories coming from the six-dimensional (2,0) theory, see \cite{Witten:2009at}.) However, there is no known description of these theories in terms of an action functional.

Above examples show that there are still many open problems in finding an appropriate action of
(relativistic or non-relativistic) gravity theories. This motivated us to employ a systematic way to
construct a non-Abelian gauge theory which is also invariant under general
coordinate transformations, in pursuit of a simple description of an (effective) action invariant under two sets of local symmetries.

With this respect, it is well-known that gravity can be obtained as a Poincar\'{e} gauge theory, where the fundamental field belongs to the representation of the Poincar\'{e} gauge group (see, for example, the textbook \cite{Blagojevic:2002du}, and the references therein). In particular, when the spacetime is Riemannean, the action becomes the Palatini one \cite{Palatini,Tsamparlis,Ferraris}, invariant under Lorentz transformations and spacetime diffeomorphisms.
In these cases, the theories are diffeomorphism invariant so that the non-Abelian gauge fields describe gravitational degrees of freedom, while the matter fields are additional ingredients, and are not part of the same gauge connection.

One of our aims is to provide a symmetry-based argument to include gravitational interactions and matter-gravity interactions in
a theory. The models which have already implemented that description are Chern-Simons gravity and supergravity
for de Sitter, anti-de Sitter or Poincar\'{e} gauge groups in three \cite{Witten:1988hc,Achucarro:1987vz} and any other odd
spacetime dimension \cite{Chamseddine:1990gk,Chamseddine:1989nu}. Inclusion of the new bosonic and fermionic fields in these theories becomes
simply an extension of the gauge group (for example to higher-spin group or
super group), at the same time fixing the interaction in an almost unique
way, using only the group theory arguments.

In this work, we use the method based on Hamiltonian formalism \cite{Henneaux-Teitelboim}, where the basic property of the diffeomorphism-invariant theories is that their Hamiltonian vanishes on-shell. It means that the diffeomorphism-invariant Hamiltonian becomes a linear combination of all its constraints that generate local symmetries in the theory. The opposite statement, in general, is not true -- the vanishing Hamiltonian does not necessarily imply that the theory is invariant under general coordinate transformations.

In a standard approach, one starts from a given action, performs a Legendre transformation in order to obtain the canonical Hamiltonian of the theory, and explores its symmetries using the Dirac method \cite{Dirac} (for a review, see for example \cite{Blagojevic:2002du,Henneaux-Teitelboim}). It results in obtaining an extended Hamiltonian as a linear combination of all symmetry-generating constraints of canonical fields present in the theory.

In our approach, we will invert the Dirac's procedure, and construct an extended Hamiltonian
which is a linear combination of our desired symmetry generators (spacetime
diffeomorphisms and non-Abelian gauge symmetry). Having the Hamiltonian, we will seek for an inverse
Legendre transformation that will give us a corresponding canonical action.
A result is an effective action of the theory invariant under given local symmetries.

The main challenge of our method is to find a canonical representation of
the constraints in terms of the canonical variables, a connection $A_{\mu}^a (x)$ which transforms in the adjoint representation of the gauge group and the corresponding canonical momenta $\pi^{\mu}_a (x)$. With respect to spacetime diffeomorphisms, in principle one could be able to construct a representation of the so-called hypersurface deformation algebra, or Dirac algebra \cite{Hojman,Thiemann}, that generates diffeomorphisms on-shell. However,  since the Dirac algebra is not a Lie algebra, it turns out that the representation of the generator of time-like diffeomorphisms is particulary intricate. For that reason, as the first step, in this work we will focus solely on the canonical representations of the generators of gauge transformations and spatial diffeomorphisms, not requiring invariance under the time reparametrizations. The actions obtained in that way are suitable for description of non-relativistic models, for example.

Let us emphasize that, when the action is known, it is straightforward to find
canonical representations of generators using the Dirac's method. Such examples which are diffeomorphism and gauge
invariant in any dimension $D$ are topological BF theories \cite{Horowitz:1989,Friedel:1999}, whose fundamental fields are the connection (with the field strength $F$) and the field $B$ which is the $D-2$ form.
Under some additional restrictions BF  theories describe gravity in first order formalism, either in complex self-dual
Ashtekar connection or the real Ashtekar-Barbero connection \cite{Ashtekar-Lewandowski,Freidel&Speziale,Durka}.
Nevertheless, even though these examples use
non-Abelian connection, they possess additional fields and, in order to obtain gravity, the original symmetry of the BF theory must be broken when $D>3$.
In two dimensions, the approach based on Ashtekar-Barbero variables can be applied for dilaton gravity and leads to the corresponding canonical representation of symmetry and spacetime diffeomorphisms generators \cite{CKRV}.

Another relevant issue is related to the uniqueness of the representation, for a given choice of canonical variables.  One of the important results is given in Ref.\cite{Hojman}, where the unique canonical representation of the hypersurface deformation algebra in four dimensions has been constructed (up to a canonical transformation of momenta), in the case when the canonical variables are the spatial 3-metric and its conjugate momentum. Furthermore, it coincides with the ADM expressions for the constraints \cite{ADM}.

The choice of the phase space variables is crucial for our analysis. For example, the gauge theories in curved background spacetimes are not diffeomorphism invariant, but can be made invariant by including the embedding variables to the configuration space \cite{Isham:1985}. The resulting theory on this extended phase space is called a parametrised field theory and is equivalent to the original one only if additional constraints are imposed. The parametrised theories are gauge and diffeomorphism invariant. Other extensions of the phase space which enable diffeomorphism invariance are possible, as for the parametrised Maxwell theory, but as a result the gauge symmetry is broken \cite{Kuchar-Stone}.

The method adopted in this work was first applied successfully in two dimensions to obtain the
gauged Wess-Zumino-Witten action \cite{Sazdovic} as a Lagrangian realization
of the Virasoro algebra that its constraints satisfy, and its supersymmetric version \cite{Miskovic-Sazdovic}. Similarly, the Liouville theory was obtained as a
gravitational Wess-Zumino action of the Polyakov string in \cite{Popovic:2000zn}, and an action for the spinning string was found in
Refs.\cite{Brink:1976sz,Brink:1976sc,Deser:1976rb}. First attempts to apply the method
in three dimensions are presented in Ref.\cite{Miskovic-Vukasinac}. The last
work was inspired by Hamiltonian analysis of Chern-Simons theories performed
in higher dimensions \cite{Banados:1996yj}.

It is worthwhile noticing that the theories described by canonical actions will be first order by construction.

In what follows, we focus on three-dimensional spacetimes. In Section \ref{P}, we define the method in a  precise way and list our assumptions.
In Section \ref{Can}, we construct symmetry generators in terms of canonical variables for both gauge generators and spatial diffeomorphisms.
In Section \ref{CanA}, we find canonical actions in three different settings, one of them reproducing the well-known Chern Simons gravity. We summarize our results in Section \ref{Dis}, pointing out open problems and possible future lines of research.

\section{Preliminaries} \label{P}

Consider a theory that is gauge-invariant under
transformations of a semi-simple Lie group $G$. For simplicity, we consider a three dimensional spacetime,
parametrized by the local coordinates $x^{\mu }=(x^0,x^i)$, $i=1,2$, of the
topology $\mathbb{R}\times \Sigma $ and the signature $(-,+,+)$, where $x^0=t$ is the time coordinate.

Let $T_{a}$, $a=1,\ldots ,n$, be anti-Hermitean generators of the corresponding Lie algebra,
\begin{equation}
[T_{a},T_{b}]=f_{ab}^{\ c}\,T_{c}\,,\qquad g_{ab}=i\mathrm{Tr}(T_{a}T_{b})\,,  \label{algebra0}
\end{equation}
where $f_{ab}^{\ c}$ are the structure constants and $g_{ab}$ is the
non-degenerate Cartan-Killing metric. The basic field is a connection $A_{\mu}^{a}(x)$
which transforms in the adjoint representations of the gauge
group, such that\ the covariant derivative of a vector field reads $D_{\mu
}V^{a}=\partial _{\mu }V^{a}+f_{bc}^{a}\,A_{\mu }^{b}V^{c}$ and the
corresponding field-strength is given by the standard expression, $F_{\mu
\nu }^{a}=\partial _{\mu }A_{\nu }^{a}-\partial _{\nu }A_{\mu
}^{a}+f_{bc}^{\ a}A_{\mu }^{b}A_{\nu }^{c}$.

In the phase space, the fundamental fields are conjugated variables $A_{\mu}^a(x)$ and the corresponding
canonical momenta $\pi_a^{\mu}(x)$,  whose Poisson bracket is given by
\begin{equation}
\{A_{\mu }^{a}(t,\vec{x}),\pi _{b}^{\nu }(t,\vec{x}^{\prime })\}=\delta
_{b}^{a}\delta _{\nu }^{\mu }\delta ^{(2)}(\vec{x}-\vec{x}^{\prime })\,.
\end{equation}
From now on, we will use the short-hand notation $\{A_{\mu }^{a},\pi_{b}^{\prime \nu }\}=\delta _{b}^{a}\delta _{\nu }^{\mu }\delta $.

In the canonical formalism, the dynamics of the theory is governed by the Hamiltonian density $H(A,\pi ,u)$,
that can also depend on the arbitrary multipliers $u(x)$. An alternative way to describe the dynamics is in terms of
the canonical action, obtained as a Legendre transformation of the Hamiltonian density,
\begin{equation}
I[A,\pi ,u]=\int dt\,d^{2}x\,[\dot{A}_{\mu }^{a}\pi _{a}^{\mu }-H(A,\pi
,u)]\,.  \label{action}
\end{equation}

In Dirac formalism, the systems with local symmetries have constraints, or on-shell
vanishing functions of canonical variables. There are two types of them:
first class constraints that generate gauge symmetries, and second class
constraints that do not generate any symmetry, but they eliminate
redundant degrees of freedom.

Keeping that in mind, we assume that our theory (\ref{action}) fulfills the
following conditions, which ensure that it is both gauge and diffeomorphism
invariant:

\begin{itemize}
\item[(\textit{i})] The theory is invariant under the action of the gauge group $G$,
so there are $2n$ independent first class constraints $G_{a}(A,\pi )=0$ and
$\bar{G}_{a}(A,\pi )=0$ generating a local symmetry with the parameters $%
\lambda ^{a}(x)$ and $\varepsilon ^{a}(x)$, respectively.

\item[(\textit{ii})] It is also invariant under general coordinate
transformations (spacetime diffeomorphisms), generated by first class
constraints $\mathcal{H}_{\mu }(A,\pi )=0$, with associated local parameter
$\xi ^{\mu }(x)$.

\item[(\textit{iii})] A corresponding Hamiltonian density is a pure constraint,
$H(A,\pi ,u)=0$, as a consequence of reparametrization symmetry.

\item[(\textit{iv})] There is an even number of second class constraints,
$\phi _{M}(A,\pi )=0$, $M=1,\ldots ,2m$ ($m\geq 0$). The functions $\phi _{M}$ must be gauge and diffeomorphism covariant.
\end{itemize}

To justify the condition (\textit{i}), recall that
first class constraints are related to local symmetries through  some local parameter
$\lambda ^{a}(x)$ and its time derivatives. However, in the Hamiltonian
formalism, all time derivatives of parameters are treated as independent
parameters.
Since our aim is to obtain a Lagrangian whose equations of motion contain at most second time derivatives, the parameter of gauge transformations depends at most on first time derivatives.
Thus, each time when  there is a local symmetry with a parameter $\lambda^{a}$, there is also
a local symmetry with a parameter $\varepsilon ^{a}\sim \dot{\lambda}^{a}$.
In consequence, there are two sets of first class constraints \cite{Castellani:1981us}, as expressed in the
condition (\textit{i}). If the gauge transformations are linear in time
derivatives, $\lambda ^{a}$ and $\varepsilon ^{a}$ are the complete set of
gauge parameters.

We also assume that the constraints are independent. This requirement is known as regularity condition
(see, for example, Ref.\cite{Henneaux-Teitelboim}). So-called irregular systems
do not fulfill this condition and the dependence of constraints produces a change in the
number of symmetries and degrees of freedom \cite{Chandia:1998uf,Miskovic:2003ex}.

The condition (\textit{ii}) ensures first class constraints which generate spacetime diffeomorphisms. As discussed in details in
Refs.\cite{Thiemann,Ashtekar-Lewandowski}, it is still an open question how to construct a canonical representation of generators of time-like diffeomorphisms on the full phase space and off-shell. For that reason, we shall focus on a representation of the spatial diffeomorphisms and construct a theory that is invariant under the transformations generated by them. The spatial diffeomorphisms can be either linearly independent or dependent on gauge symmetries. For example, all spatial diffeomorphisms are independent in higher-dimensional Chern-Simons theory, and dependent on gauge transformations in three-dimensional spacetime \cite{Banados:1996yj}. Thus, to have a theory covariant on $\Sigma $, the spatial diffeomorphisms are
either \textit{all} dependent or \textit{all} independent on the gauge
symmetries, while the time-like diffeomorphisms, that generate time evolution off $\Sigma$, can be treated separately.
Therefore, the number of independent constraints is $2\varepsilon $ for spatial
diffeomorphisms and $\varepsilon _{0}$ for time-like diffeomorphism, where
$\varepsilon ,\varepsilon _{0}=0$ or $1$.

In general case, the representation of the generator of the time-like
diffeomorphism $\mathcal{H}_{0}$, when it is independent on gauge
transformations, is not unique. In order to avoid technical difficulties
related to its representation, we shall not include it. It means we will set
$\varepsilon _{0}=0$, which can imply that the obtained theory is not invariant under
the time-like diffeomorphism.
Then one can ask whether the extended Hamiltonian still vanishes on-shell. As pointed out in \cite{Henneaux-Teitelboim}, the answer is -- not necessarily, one example is an (effective) theory invariant under spatial diffeomorphisms with the canonical Hamiltonian different than zero \cite{Husain:2011tk}. However, since our ultimate goal is obtention of a fully covariant theory, we will not consider these cases.
On the contrary, when possible, we can try to covariantize the final action, so that it becomes a scalar density under the general
coordinate transformations, expecting that the Hamiltonian constraint would be implemented in that way dynamically.
In other cases, when the covariantization is not possible, the new action should be seen as the one with reduced symmetry, describing in that way theories such as non-relativistic ones, or the ones in presence of a membrane.

The condition (\textit{iii}) is a property of a diffeomorphism invariant theory. Note that here and throughout this manuscript the sign of equality refers also to on-shell equality.

As for the last condition (\textit{iv}), we permit the existence of second
class constraints, whose number is always even. In order to perform the
inverse Legendre transformations, when it is possible, we can only use the
\textit{second class} constraints, definitions of velocities $\dot{A}(A,\pi)$ and values for $u$ obtained from
the evolution of the constraints. We cannot use all equations of motion because then the action vanishes on-shell,
as it becomes a boundary term \cite{Hojman-Urrutia}.

At the end, let us count the degrees of freedom of our system. We start with
$3n$ Lagrangian fields $A_{\mu }^{a}$, and have $2n+2\varepsilon$
first class constraints ($G_{a},\bar{G}_{a},\mathcal{H}_{i }$) and $2m$ second class constraints $\phi _{M}$. A number of degrees
of freedom $N$ in the theory is, therefore, $3n-(2n+2\varepsilon )
-\frac{1}{2}\,2m$, or
\begin{equation}
N=n-m-2\varepsilon .  \label{deg}
\end{equation}

In the following section we will construct a representation of the
constraints, as a first step towards the formulation of the canonical action.

\section{Canonical representation of the generators} \label{Can}

We seek for a theory invariant under internal gauge transformations and spatial
diffeomorphisms. To this end, we need a representation of the generators in
terms of canonical variables $\left( A,\pi \right) $. This representation is
not, in general, unique, and we will choose a particular one and
discuss possible generalizations.

\subsection{Gauge generators}

As mentioned before, the canonical gauge generators $G_{a}(A,\pi )$ are
first class constraints and they satisfy the Lie algebra
\begin{equation}
\left\{ G_{a},G_{b}^{\prime }\right\} =f_{ab}^{\ c}\,G_{c}\,\delta \,.
\label{algebra1}
\end{equation}%
Remind that we are using the short-hand notation, such that $G_b^{\prime} =G_b (t, \vec{x}^{\prime})$.
The algebra of the generators $\bar{G}_{a}(A,\pi )$ does not have predetermined
form, and has to be such that they generate, together with $G_{a}$, via the
Poisson brackets, desired gauge transformations%
\begin{equation}
\delta A_{\mu }^{a}=\{A_{\mu }^{a},G[\lambda ,\varepsilon ]\}=-D_{\mu
}\lambda ^{a}\,,  \label{DLambda}
\end{equation}%
using the total, smeared symmetry generator
\begin{equation}\label{GeneratorGauge}
{\mathcal G}[\lambda ,\varepsilon ]=\int d^{2}x\,\left( \lambda ^{a}G_{a}+\varepsilon
^{a}\bar{G}_{a}\right) \,.
\end{equation}

It is worthwhile to emphasize that the smeared generator, as well as the
Hamiltonian and all functionals of canonical variables, have to be
differentiable quantities in order to have well-defined Poisson brackets.
This is achieved by supplementing them with suitable boundary terms. This
step is essential also for definition of canonical conserved charges
\cite{Regge-Teitelboim,Banados-charges}. In this text, however, we will
not discuss the differentiability of the functionals as we are, at the
initial stage, concerned with the construction of the canonical action
possessing given symmetries. The question of the boundary terms should be,
however, addressed later on. {From} now on, we will neglect all
boundary terms in our calculations.

Recall that we introduced the additional set of generators $\bar{G}_{a}$ in
order to deal, in a canonical way, with the time derivatives of the local
parameters that appear in the gauge transformation.
%
Namely, since $\dot{%
\lambda}^{a}$ becomes an independent parameter represented by $\varepsilon
^{a}$, the  relation between them is naturally induced by the gauge
transformation of the form (\ref{DLambda}), namely,
\begin{eqnarray}
\delta A_{i}^{a} &=&\{A_{i}^{a},{\mathcal G}[\lambda ,\varepsilon ]\}=-D_{i}\lambda
^{a}\,,  \notag \\
\delta A_{0}^{a} &=&\{A_{0}^{a},{\mathcal G}[\lambda ,\varepsilon ]\}=-D_{0}\lambda
^{a}\equiv \varepsilon ^{a}\,.  \label{DLambda,Epsilon}
\end{eqnarray}
Therefore, $G_{a}$ and $\bar{G}_{a}$ have to be such that the above
transformation law is satisfied.

Let us start with the transformation law of the spatial components
$A_{i}^{a} $. The first line in (\ref{DLambda,Epsilon}) and independence of
$\lambda ^{b}$ and $\varepsilon ^{b}$ lead to the following functional
equations in $G_{a}$ and $\bar{G}_{a}$,
\begin{eqnarray}
\int d^{2}x^{\prime }\,\lambda ^{\prime b}\{A_{i}^{a},G_{b}^{\prime }\}
&=&-D_{i}\lambda ^{a}\,,  \notag \\
\int d^{2}x^{\prime }\,\varepsilon ^{\prime b}\{A_{i}^{a},\bar{G}_{b}^{\prime }\} &=&0\,,
\end{eqnarray}
or equivalently
\begin{equation}
\frac{\partial G'_{b}}{\partial \pi _{a}^{i}}=\delta_{b}^{a}\,
\partial^{\prime}_{i}\delta +f_{bc}^{\ \ a}\,A_{i}^{\prime c}\delta \,,
\qquad \frac{\partial \bar{G}_{b}^{\prime }
}{\partial \pi _{a}^{i}}=0\,. \label{gauge eqs}
\end{equation}
Note that, since $G_{a}$ and $\bar{G}_{a}$ can depend at most on first
spatial derivatives of the gauge field, the quantities of the type $\frac{
\partial G_{a}}{\partial A_{i}^{\prime b}}$ or $\frac{
\partial G_{a}}{\partial \pi^{\prime i}_{b}}$
are distributions (Dirac delta function and its first spatial derivatives).

The general solution of Eqs.(\ref{gauge eqs}) reads
\begin{eqnarray}
G_{a} &=&D_{i}\pi _{a}^{i}+h_{a}(A_{\mu }^{b},\pi _{b}^{0})\,,  \notag \\
\bar{G}_{a} &=&\bar{G}_{a}(A_{\mu }^{b},\pi _{b}^{0})\,,  \label{G,barG}
\end{eqnarray}
where $h_{a}$ and $\bar{G}_{a}$ are arbitrary phase space functions
independent on $\pi _{a}^{i}$. In the special case when $h_a=0$, $G_a$ reduces to the Gauss constraint, which is the analogue of Gauss’ law of electromagnetism with the momentum being an electric field. Eqs.(\ref{G,barG}) are its generalization to the non-Abelian case.

On the other hand, from the transformation law of the time-like component
$A_{0}^{a}$, it follows that%
\begin{eqnarray}
\int d^{2}x^{\prime }\,\lambda ^{\prime b}\{A_{0}^{a},G_{b}^{\prime }\}
&=&0\,,  \notag \\
\int d^{2}x^{\prime }\,\varepsilon ^{\prime b}\{A_{0}^{a},\bar{G}_{b}^{\prime }\} &=&\varepsilon ^{a}\,,
\end{eqnarray}
or equivalently
\begin{equation}
\frac{\partial h_{b}^{\prime }}{\partial \pi _{a}^{0}}=0\,,\qquad
\frac{\partial \bar{G}_{b}^{\prime }}{\partial \pi _{a}^{0}}=\delta _{b}^{a}\delta
\,,
\end{equation}
leading to the solution
\begin{eqnarray}
h_{a} &=&h_{a}(A_{\mu }^{b})\,,  \notag \\
\bar{G}_{a} &=&\pi _{a}^{0}+\bar{h}_{a}(A_{\mu }^{b})\,.  \label{F,barF}
\end{eqnarray}%
We conclude that both first class constraints $G_{a}$ and $\bar{G}_{a}$ are
linear in momenta in case of non-Abelian gauge symmetries.

The algebra of the gauge generators given by the expressions (\ref{G,barG}) and (\ref{F,barF}) has the following form,
\begin{eqnarray}
\{G_{a},G_{b}^{\prime }\} &=&f_{ab}^{\ \ c}G_{c}\,\delta +\Delta
G_{ab}^{(1)}\,,  \notag \\
\{G_{a},\bar{G}_{b}^{\prime }\} &=&\Delta G_{ab}^{(2)}\,,  \notag \\
\{\bar{G}_{a},\bar{G}_{b}^{\prime }\} &=&\Delta G_{ab}^{(3)}\,,
\end{eqnarray}
where
\begin{eqnarray}
\Delta G_{ab}^{(1)}(x,x^{\prime }) &=&-f_{ab}^{\ \ c}h_{c}\,\delta +f_{bc}^{\
\ d}A_{i}^{c}\,\frac{\partial h_{d}^{\prime }}{\partial A_{i}^{a}}-f_{ac}^{\
\ d}A_{i}^{\prime c}\,\frac{\partial h_{d}}{\partial A_{i}^{\prime b}}%
+D_{i}^{\prime }\left( \frac{\partial h_{a}}{\partial A_{i}^{\prime b}}%
\right) -D_{i}\left( \frac{\partial h_{b}^{\prime }}{\partial A_{i}^{a}}%
\right) \,,  \notag \\
\Delta G_{ab}^{(2)}(x,x^{\prime }) &=&-D_{i}\left( \frac{\partial \bar{h}%
_{b}^{\prime }}{\partial A_{i}^{a}}\right) +f_{bc}^{\ \ d}A_{i}^{c}\,\frac{%
\partial \bar{h}_{d}^{\prime }}{\partial A_{i}^{a}}+\frac{\partial h_{a}}{%
\partial A_{0}^{\prime b}}\,,  \notag \\
\Delta G_{ab}^{(3)}(x,x^{\prime }) &=&-\frac{\partial \bar{h}_{b}^{\prime }}{%
\partial A_{0}^{a}}+\frac{\partial \bar{h}_{a}}{\partial A_{0}^{\prime b}}\,.
\label{DeltaG}
\end{eqnarray}

In order to describe a considered non-Abelian symmetry, the algebra has to close without addition of new constraints, so the right hand side should be either equal to zero or proportional to the constraints (\ref{G,barG}) linear in momenta. It implies that, because $h_a$ and $\bar{h}_a$ do not depend on the momenta, all extra terms depending on these functions on the right hand side of Eqs.(\ref{DeltaG}) have to vanish.

The result are the restrictions
\begin{equation}
\Delta G_{ab}^{(n)}=0\,,\qquad n=1,2,3. \label{conditions}
\end{equation}

One particular family of two-parameter solutions of these restrictions is
\begin{equation}
h_{a}=\alpha \,g_{ab}\varepsilon ^{ij}\partial _{i}A_{j}^{b}\,,\qquad \bar{h}%
_{a}=\beta \, g_{ab} A_0^b\,,  \label{h=dA}
\end{equation}%
where $\alpha $ and $\beta $ are real constants, which can also be zero.
Indeed, it is straightforward to show that all covariant derivatives appearing
in $\Delta G_{ab}^{(n)}$ vanish individually, for example $D_{i}\left( \frac{%
\partial h_{b}^{\prime }}{\partial A_{i}^{a}}\right) =0$, and similarly for
other covariant derivatives. Using also the facts that $h_{a}$ and $\bar{h}%
_{a}$ are independent on $A_{0}^{b}$ and $A_{i}^{b}$, respectively, the
other terms cancel out, for example
\begin{equation}
f_{bc}^{\ \ d}A_{i}^{c}\,\frac{\partial h_{d}^{\prime }}{\partial A_{i}^{a}}%
-f_{ac}^{\ \ d}A_{i}^{\prime c}\,\frac{\partial h_{d}}{\partial
A_{i}^{\prime b}}=-\alpha f_{ab}^{\ \ c}\,\varepsilon ^{ij}g_{cd}\left(
A_{i}^{d}\,\partial _{j}^{\prime }\delta +A_{i}^{\prime d}\,\partial
_{j}\delta \right) =f_{ab}^{\ \ c}\,h_{c}\,\delta \,.
\end{equation}%
In that way, the choice (\ref{h=dA}) yields $\Delta G_{ab}^{(n)}(x,x^{\prime
})=0$ for all $n$ and the non-Abelian gauge algebra closes.

The algebra of the constraints becomes
\begin{eqnarray}
\{G_{a},\bar{G}_{b}^{\prime }\} &=&0\,,\qquad \qquad \{\bar{G}_{a},\bar{G}%
_{b}^{\prime }\}=0\,,  \notag \\
\{G_{a},G_{b}^{\prime }\} &=&f_{ab}^{c}G_{c}\delta \,,  \label{GG}
\end{eqnarray}%
and the generator of gauge transformations acquires the form
\begin{equation}
{\mathcal G}[\lambda ,\varepsilon ]= \int d^{2}x\, \left[ \lambda^a
( D_{i}\pi _{a}^{i} + h_a(A)) +\varepsilon^{a}(\pi_a^0 + \bar{h}_a(A)  ) \right],
\end{equation}
where $h_a(A)$ and $\bar{h}_a(A)$ satisfy the conditions (\ref{conditions}).

\subsection{Diffeomorphism generators}

Now we focus on the representation of the generator of general coordinate
transformations, $\delta x^{\mu }=-\xi ^{\mu }(x)$, that acts on the fields
as a Lie derivative, $\delta A_{\mu }^{a}=\pounds_{\xi }A_{\mu }^{a}$. A canonical representation of the generators of these transformations, spacetime diffeomorphisms, should be such that the smeared generator $D[\xi ]=\int d^2x\, \xi^{\mu} \mathcal{H}_{\mu}$ acts on an arbitrary function of phase space variables $F(A,\pi )$ as
\begin{equation}
\delta_\xi F(A,\pi ) = \{ F(A,\pi ),  D[\xi ] \} =   F(\pounds_{\xi }A, \pounds_{\xi}\pi )\, .
\end{equation}

The diffeomorphism group represents a \textit{kinematical} symmetry of any diffeomorphism invariant
action. In the standard approach, one starts from the Lagrangian of a diffeomorphism invariant theory and
constructs the canonical representation of the generators of spacetime diffeomorphisms. The canonical representation of generators of space-like diffeomorphisms, $\mathcal{H}_{i}(A,\pi )$, is independent on the Lagrangian, while the generator of time-like diffeomorphism, so called the Hamiltonian constraint, $\mathcal{H}_{0}(A,\pi )$, depends on the dynamics of the theory (the form of the Lagrangian). For that reason, they generate  a \textit{dynamical} symmetry of the theory. The fact that the dynamics takes  place on spacelike hypersurfaces embedded in a spacetime with Lorentzian signature  is reflected in the algebra of the constraints $(\mathcal{H}_{0}, \mathcal{H}_{i})$,
which is Dirac algebra, also known as hypersurface deformation algebra, of the form \cite{Dirac,Kuchar-Stone}
\begin{eqnarray}
\{  \mathcal{H}_{0 },\mathcal{H}_{0 }^{\prime}\} &=& (q^{ij}\mathcal{H}_{j}+q^{\prime ij}\mathcal{H}_{j}^{\prime})\partial_i\delta\, , \label{00}\\
\{  \mathcal{H}_{i },\mathcal{H}_{0 }^{\prime}\} &=& \mathcal{H}_{0 }\partial_i\delta\, ,\label{0i}\\
\{  \mathcal{H}_{i },\mathcal{H}_{j }^{\prime}\} &=& \mathcal{H}_{i }^{\prime}\partial_j\delta + \mathcal{H}_{j }\partial_i\delta\, , \label{ij}
\end{eqnarray}
where $q^{ij}$ is the inverse of the induced metric on the spatial hypersurface. The Dirac algebra is not a Lie algebra, because of the structure functions $q^{ij}(x)$ in the brackets of the Hamiltonian constraints (\ref{00}). For that reason,
the canonical representations of the Hamiltonian constraint and {the} spatial diffeomorphism {generator} cannot be derived in the same way.

In particular, the time evolution off $\Sigma$ is determined by the Hamiltonian constraint $\mathcal{H}_{0}$, so its representation is theory-dependent and cannot be obtained in a straightforward way by our methods. As explained in the previous section (see discussion about the point (\textit{ii})), we shall focus only on the spatial diffeomorphisms.
 Under their action, the field $A_{0}^{a}$ transforms as a scalar and the field $A_{i}^{a}$ transforms as a one-form. Similarly, $\pi _{a}^{0}$ and $\pi_{a}^{i}$ are scalar and vector densities, respectively.

Now we can make use {of} the fact that, apart from general coordinate transformations, we also have gauge symmetry. Recall that the Cartan's identity,
\begin{equation}
\pounds _{\xi }A_{\mu }^{a}=D_{\mu }\left( \xi ^{\nu }A_{\nu }^{a}\right)
-\xi ^{\nu }F_{\mu \nu }^{a}\,,  \label{diff-gauge identity}
\end{equation}%
relates the field-dependent gauge transformation $D_{\mu }\left( \xi ^{\nu
}A_{\nu }^{a}\right) $ to the diffeomorphisms $\pounds _{\xi }A_{\mu }^{a}$.
In the special case of theories where the equations of motion are $F_{\mu
\nu }^{a}=0$, these two sets of local transformations are dependent
on-shell, $\pounds _{\xi }A_{\mu }^{a}=D_{\mu }\lambda ^{a}$. Clearly, in
these theories $A_{\mu }^{a}$ is a pure-gauge, so the theory has no local
degrees of freedom, $N=0$.

Local symmetry represented by the second term on the r.h.s.~of Eq.(\ref{diff-gauge identity}), relevant only when $N\neq 0$, is also called improved diffeomorphism \cite{Jackiw:1978ar}.

The identity (\ref{diff-gauge identity}) applied to the spatial
diffeomorphisms suggests the form of the spatial diffeomorphism generators
as
\begin{eqnarray}
\mathcal{H}_{i}^{\mathrm{aux}} &=&-A_{i}^{b}G_{b}+F_{ik}^{b}\pi _{b}^{k}
\notag \\
&=&-A_{i}^{b}\partial _{k}\pi _{b}^{k}+(\partial _{i}A_{k}^{b}-\partial
_{k}A_{i}^{b})\pi _{b}^{k}-A_{i}^{b}h_{b}\,,  \label{gen_spatial_diffeo aux}
\end{eqnarray}%
so that the smeared generator would read
\begin{equation}
\mathcal{H}^{\mathrm{aux}}[\xi ]=\int d^{2}x\,\xi ^{i}\mathcal{H}_{i}^{%
\mathrm{aux}}\,.
\end{equation}%
In that case, it is straightforward to check that, under spatial
diffeomorphisms, $A_{i}^{a}$ transforms like a one-form, for any function $h_{b} $,
\begin{equation}
\{A_{i}^{a},\mathcal{H}^{\mathrm{aux}}[\xi ]\}=\partial _{i}\xi
^{j}\,A_{j}^{a}+\xi ^{j}\partial _{j}A_{i}^{a}=\pounds _{\xi }A_{i}^{a}\,,
\end{equation}%
while $\pi _{a}^{i}$ should change as a vector density of the weight one $\pounds %
_{\xi }\pi _{a}^{i}=\partial _{j}(\xi ^{j}\pi _{a}^{i})-(\partial _{j}\xi^{i})\,\pi _{a}^{j}$. Straightforward calculation gives
\begin{equation}
\{\pi _{a}^{i},\mathcal{H}^{\mathrm{aux}}[\xi ]\}=\pounds _{\xi }\pi
_{a}^{i}+\int d^{2}x^{\prime }\,\xi ^{\prime j}\frac{\partial }{\partial
A_{i}^{a}}\left( A_{j}^{\prime b}h_{b}^{\prime }\right) \,,
\end{equation}%
implying that the last term in the above formula has to vanish (on-shell or off-shell). However, if we
exclude the term $-A_{i}^{b}h_{b}$ from the diffeomorphism generator, we
obtain a good transformation law of the fields in any case, and
independently on the function $h_{a}$. We shall, therefore, define the
spatial diffeomorphisms generator as
\begin{equation}
\mathcal{H}[\xi ]=\int d^{2}x\,\xi ^{i}\mathcal{H}_{i}\,,
\end{equation}%
where $\mathcal{H}_{i}$ is linear in $A_{i}^{a}$ and $\pi _{a}^{i}$,%
\begin{eqnarray}
\mathcal{H}_{i} &=&-A_{i}^{b}D_{k}\pi _{b}^{k}+F_{ik}^{b}\pi _{b}^{k}  \notag
\\
&=&-A_{i}^{b}\partial _{k}\pi _{b}^{k}+(\partial _{i}A_{k}^{b}-\partial
_{k}A_{i}^{b})\pi _{b}^{k}\,.  \label{gen_spatial_diffeo}
\end{eqnarray}%
Then, $\delta_{\xi}F(A,\pi )=\{F(A,\pi ),\mathcal{H}[\xi ]\}=F(\pounds_{\xi }A, \pounds_{\xi}\pi )$,
is satisfied for any function on the phase space.

Direct {checkup} shows that $\mathcal{H}_{i}$ close the spatial
diffeomorphisms sub-algebra (\ref{ij}),
but the algebra with gauge generators $\{G_{a},\mathcal{H}_{i}^{\prime }\}$ does not close, when $h_a(A)\ne 0$, even if the conditions (\ref{conditions}) are fulfilled. For that reason, {for now} we shall restrict our analysis to the case when $h_{a}=0$ and $\bar{h}_{a}=0$, {leading to}
\begin{eqnarray}
\{G_{a},\mathcal{H}_{i}^{\prime }\} &=&G_{a}^{\prime }\partial _{i}\delta \,,
\notag \\
\{\bar{G}_{a},\mathcal{H}_{i}^{\prime }\} &=&0\,.  \label{GH}
\end{eqnarray}
The cases $h_{a},\bar{h}_{a}\neq 0$ will become relevant when the second class constrains are included, and will be discussed in the next section.

Furthermore, it is more convenient to replace $\mathcal{H}_{i}$ by other first class constraints $K_{i}$,
\begin{equation}\label{K}
K_{i}\equiv F_{ik}^{b}\pi _{b}^{k}=\mathcal{H}_{i}+A_{i}^{b}G_{b}\,.
\end{equation}%
It turns out that $K_{i}$  are generators of improved spatial diffeomorphisms \cite{Jackiw:1978ar}, which differ from spatial diffeomorphisms by a gauge transformation.  Namely, if we define a smeared generator
\begin{equation}
K[\xi ]=\int d^{2}x\,\xi ^{j}K_j\, ,
\end{equation}
then the improved diffeomorphisms transform the fields as
\begin{equation}
\{A_{i}^{a}, K[\xi ]\} = \xi^j F_{ji}^a =  \pounds_{\xi }A_{i}^{a} - D_i (\xi^j A_j^a)\, .
\end{equation}
New constraints satisfy the following algebra,
\begin{eqnarray}
\{G_{a},K_{i}^{\prime }\} &=&0\,,  \notag \\
\{K_{i},K_{j}^{\prime }\} &=&K_{i}^{\prime }\partial _{j}\delta
+K_{j}\partial _{i}\delta -F_{ij}^{a}G_{a}\delta \,.  \label{GK}
\end{eqnarray}
Note that the Poisson brackets of $K_{i}$ does not close without the presence of $G_{a}$ and there are field-dependent structure functions $F_{ij}^{a}$.

Remarkably, in the Hamiltonian analysis of the higher-dimensional Chern-Simons theory, the constraints of the improved diffeomorphisms appear naturally as secondary first  class constraints \cite{Banados:1996yj}, and the diffeomorphisms are only an on-shell symmetry. This example suggests that it is simpler to choose $K_{i}$ as the symmetry generators for construction of the Hamiltonian action in general.

To summarize, the constraints in the theory invariant under both gauge
transformations and spatial diffeomorphisms include the generators of these
symmetries, but it can also contain a set of second class constraints that
are not symmetry generators, {and whose presence is} important for building a covariant
theory. Therefore, the complete set of the constraints is:
\begin{equation*}
\fbox{$%
\begin{array}{lll}
\mathrm{First\;class:} & \bar{G}_{a}=\pi _{a}^{0}\,, & n\,,
\\
& G_{a}=D_{i}\pi _{a}^{i}\,,\qquad  & n\,, \\
& {K}_{i}=F_{ik}^{b}\pi _{b}^{k}\,, &
2\varepsilon \,,\medskip  \\
\mathrm{Second\ class:}\qquad  & \phi _{M}\,, & 2m\,.%
\end{array}
$}
\end{equation*}

\section{Canonical action} \label{CanA}

When the theory is invariant under general coordinate transformations,
the Hamiltonian becomes a pure constraint. It enables {to write}
the most general Hamiltonian density as a linear combination of all
constraints present in the theory,
\begin{equation}
H=u^{a}\bar{G}_{a}+v^{a}G_{a}+\zeta ^{i}K_{i}+w^{M}\phi _{M}\,,
\label{H}
\end{equation}
where the associated Hamiltonian multipliers are $u^{a}(x)$, $v^{a}(x)$, $%
\zeta ^{i}(x)$ and $w^{M}(x)$.

On the other hand, in the standard approach one starts from the Lagrangian description of a theory and then in the Dirac formalism distinguishes between the primary and secondary constraints,
that is, the ones obtained from the definition of the canonical momenta
(primary), and the ones obtained from the evolution of the primary
constraints (secondary).
The Legendre transformation of the Lagrangian evaluated on-shell yields
the \textit{canonical} Hamiltonian, but a proper definition of physical variables in the reduced phase space requires
also to account the constraints between canonical variables.
It leads to the \textit{total}
Hamiltonian that includes also the primary constraints, and the \textit{extended}
Hamiltonian that includes all (primary and secondary) constraints.

We will adopt an approach of Dirac who conjectured that all first class
constraints (primary and secondary) generate gauge transformations and
should be included in the extended Hamiltonian. In Ref.\cite{Castellani:1981us},
it was shown that this is always the case, except when
the powers of some constraint $\phi $ appear, e.g., $\phi^{n}$. Then the
constraint $\phi^{n}$ and the ones following from its evolution do not
enter the extended Hamiltonian. Thus, only the first class constraints that
\textit{do} generate gauge transformations are included in the extended
Hamiltonian. The resulting Hamiltonian equations of motion are not identical
to the Lagrangian ones, but the difference is not physical. For more
details, see for example Ref.\cite{Blagojevic:2002du}.

It this approach, we construct the canonical action which contains
larger number of arbitrary multipliers, obtained from the extended
Hamiltonian. These multipliers can be partially identified with unphysical
components of the gauge field if one conveniently assumes that some of the
first class constraints are primary, and therefore requires that their
evolution produces the other --secondary-- first class constraint. We will
make use of this method to reduce a number of multipliers.
Resulting equations of motion would not depend on the choice of the constraints as primary or secondary, but the effective Lagrangian could have different (physically equivalent) form, in latter case containing additional auxiliary fields \cite{Peldan}.

So far we have not considered the second class constraints, $\phi _{M}$. Recall that
there must be an even number of these constraints, because their Poisson
brackets have the form
\begin{equation}
\left\{ \phi _{M},\phi' _{N}\right\} =\Delta_{MN}(x,x')\,,\label{IIclassA}
\end{equation}
where the tensor $\Delta _{MN}$ must be invertible on-shell.
Therefore, the rank of $\Delta_{MN}$ has to be maximal, and since it is antisymmetric, it is an even number $2m$.

In addition, the constraints $ \phi _{M}$ have to commute on-shell with all first class constraints.

At first sight, it seems that there is a huge arbitrariness in the choice of
$\phi _{M}(x)$. However, although their choice is not unique, there are at
least two conditions which drastically reduce a number of possible choices.
First, their number is limited by the number of degrees of freedom in the theory and, second, they have to be
covariant under both diffeomorphisms and gauge transformations, so that the
index $M$ is not arbitrary. More precisely, its tensorial properties are
determined by the indices $a,b,\ldots$ (with the range $n$) and
$i,j,\ldots$ (with the range $2$). From the degrees of freedom count
(\ref{deg}), we have
\begin{equation}
0\leq m\leq n-2\varepsilon \,, \label{N>0}
\end{equation}
where, as discussed before, we assume $\varepsilon_0=0$. The above inequality implies, for example, that for $n>3$, $\phi _{M}$ cannot be an antisymmetric or symmetric group tensor of rank two, as $n(n-1)/2$ or $n(n+1)/2$ constraints
would not satisfy the above condition.

Keeping this argument in mind, there are the following allowed multiplets of
the constraints $\phi _{M}$ fixed by its tensorial properties,
\begin{equation*}
\fbox{$%
\begin{array}{ccccccccc}
\text{Second class constraints}: & \phi & \phi _{a} & \phi _{i} & \phi _{ai}
& \phi _{\lbrack ij]} & \phi _{(ij)} & \phi _{(ij)k} & \cdots \medskip \\
\;\;\text{Number of componets }: & 1 & n & 2 & 2n & 1 & 3 & 6 & \cdots%
\end{array}
$}
\end{equation*}
The sum of all the components present in the theory must be equal to $2m$. \medskip

So far, we found a particular representation of first class constraints and
wrote an extended Hamiltonian describing a theory with $n-m-2\varepsilon $
degrees of freedom. Let us prove now that this method indeed produces an action invariant under local transformations.

Before doing it, further simplification can be done in the Hamiltonian (\ref{H}). Namely,
since the second class constraints have to satisfy the consistency conditions,
we use Eq.(\ref{IIclassA}) and the fact that $\left\{ \phi_{M},\mathcal{G}_{A}^{\prime }\right\} =0$ on-shell, to obtain
\begin{equation}
\dot{\phi}_{M}=\left\{ \phi _{M},H\right\} =\int d^{2}x^{\prime}\,w^{\prime N}\Delta _{MN}(x,x^{\prime })=0\,.
\end{equation}
Then, due to invertibility of the matrix $\Delta _{MN}$, the multipliers associated to second class constraints
vanish, $w^{N}=0$. As a result, the extended Hamiltonian density does not {depend on} second class constraints,
{acquiring the form}
\begin{equation}
H = U^{A}\mathcal{G}_{A} \equiv u^{a}\bar{G}_{a}+v^{a}G_{a}+\zeta ^{i}K_{i}\,,
\label{H2}
\end{equation}
where $\mathcal{G}_{A}=\left( G_{a},\bar{G}_{a},K_{i}\right) $ are first class constraints and $U^{A}=\left( u^{a},v^{a},\zeta ^{i}\right) $ are
the corresponding multipliers.

Hamiltonian equations are invariant under local transformations generated by
the smeared generator
\begin{equation}
\mathcal{G}[\Lambda ]=\int d^{2}x\,\Lambda ^{A}\mathcal{G}_{A}=\int
d^{2}x\,\left( \lambda ^{a}G_{a}+\varepsilon ^{a}\bar{G}_{a}+\xi
^{i}K_{i}\right) \,,
\end{equation}
where $\Lambda ^{A}=\left( \lambda ^{a},\varepsilon ^{a},\xi ^{i}\right) $
are local parameters.
The algebra given by Eqs.(\ref{GG}) and (\ref{GK})
does not have the standard form. It closes, but with some
some structure functions  $f_{AB}^{\ \ \ C}(x)$ depending on the canonical fields.
Also, since the algebra in general involves distributions $\delta$ and $\partial_i\delta$,
the derivative term gives rise to another set of the structure functions, $f_{AB}^{\ \ \ Ci}(x)$. Thus, the Poisson brackets algebra has the form,
\begin{equation}
\left\{ \mathcal{G}_{A},\mathcal{G}_{B}^{\prime }\right\} =f_{AB}^{\ \ \ C}\,%
\mathcal{G}_{C}\,\delta +f_{AB}^{\ \ \ Ci}\,\mathcal{G}_{C}\,\partial
_{i}\delta \,,  \label{calGG1}
\end{equation}%
where the structure functions are not completely arbitrary, because the
 brackets are antisymmetric and (\ref{calGG1}) has to satisfy the Jacobi
identity.

Under these conditions, the canonical action
\begin{equation}
I=\int dtd^{2}x\,\left( \pi _{a}^{\mu }\dot{A}_{\mu }^{a}-U^{A}\mathcal{G}_{A}\right)
\end{equation}
{is} invariant under the transformation generated by {the} smeared
generator $\mathcal{G}\left[ \Lambda \right] $.
Indeed, this action changes under local transformations as
\begin{equation}
\delta I=\int dtd^{2}x\,\left( \delta \pi _{a}^{\mu }\dot{A}_{\mu }^{a}+\pi
_{a}^{\mu }\delta \dot{A}_{\mu }^{a}-\delta U^{A}\mathcal{G}_{A}-U^{A}\delta
\mathcal{G}_{A}\right) \,.
\end{equation}%
Using the algebra (\ref{calGG1}), the generators vary as
\begin{equation}
\delta \mathcal{G}_{A}=\{\mathcal{G}_{A},\mathcal{G}[\Lambda ]\}=\left(
f_{AB}^{\ \ \ C}\Lambda ^{B}+f_{AB}^{\ \ \ Ci}\partial _{i}\Lambda
^{B}\right) \mathcal{G}_{C}\,,
\end{equation}
{while} the canonical fields transform (with the functions $h_{a}$ and $\bar{h}_{a}$ given by Eqs.(\ref{h=dA})) as
\begin{eqnarray}
\delta A_{i}^{a} &=&\{A_{i}^{a},\mathcal{G}[\Lambda ]\}=-D_{i}\lambda
^{a}+\xi ^{j}F_{ji}^{a}\,,  \notag \\
\delta A_{0}^{a} &=&\{A_{0}^{a},\mathcal{G}[\Lambda ]\}=\varepsilon ^{a}\,,
\notag \\
\delta \pi _{a}^{0} &=&\{\pi _{a}^{0},\mathcal{G}[\Lambda ]\}=-\beta
\,g_{ab}\varepsilon ^{b}\,,  \notag \\
\delta \pi _{a}^{i} &=&\{\pi _{a}^{i},\mathcal{G}[\Lambda ]\}=f_{ab}^{\
d}\lambda ^{b}\pi _{d}^{i}+D_{j}\left( \xi ^{j}\pi _{a}^{i}-\xi ^{i}\pi
_{a}^{j}\right) -\alpha \,g_{ab}\,\varepsilon ^{ij}\partial _{j}\lambda
^{b}\,.
\end{eqnarray}
Plugging these expressions into $\delta I$, after few integrations by parts, first two terms in $\delta I$
lead to the identity
\begin{equation}
\int dtd^{2}x\,\left( \delta \pi _{a}^{\mu }\dot{A}_{\mu }^{a}+\pi
_{a}^{\mu }\delta \dot{A}_{\mu }^{a}\right)
=\int dtd^{2}x\,\mathcal{G}_{A}\dot{\Lambda}^{A}\,,
\end{equation}
and the full canonical action changes under the
local transformations generated by $\mathcal{G} [\Lambda ]$ as
\begin{equation}
\delta I=\int dtd^{2}x\,\mathcal{G}_{A}\left( \dot{\Lambda}^{A}
-\delta U^{A}-f_{BC}^{\ \ A}U^{B}\Lambda ^{C}-f_{BC}^{\ \ Ai}U^{B}\partial
_{i}\Lambda ^{C}\right).
\end{equation}%
The invariance ($\delta I=0$) follows if the indefinite
multipliers change according to the rule
\begin{equation}
\delta U^{A}=\dot{\Lambda}^{A}-f_{BC}^{\ \ A}U^{B}\Lambda ^{C}-f_{BC}^{\ \
Ai}U^{B}\partial _{i}\Lambda ^{C}\,.
\end{equation}
This transformation law is a generalization of the {one} given in
the review \cite{Banados-Reyes}, where the algebra {did} not contain
the $\partial _{i}\delta $ terms and whose structure functions {were}
restricted to the field-independent structure constants.

In the next section, we will choose particular constraints (for given $m$ and $\varepsilon $),
and construct the corresponding gauge theories.
Recall that we always have $\varepsilon =1$, except when $F_{\mu \nu }^{a}=0$.\medskip

Let us focus on special cases.

\subsection{Theory with the maximal number of degrees of freedom \label{NoNo}}

The simplest possible theory has the minimal number of constraints, which are
only the gauge ones, with the arbitrary functions in the equations
(\ref{G,barG}) and (\ref{F,barF}) set to zero, $h_{a}=0$, $\bar{h}_{a}=0$, and
without second class constraints ($m=0$).
Absence of spatial diffeomorphisms will either produce a theory where  all
diffeomorphisms are functionally dependent on the gauge
transformations ($\varepsilon =0$), or the theory will not be invariant under general coordinate transformations.

This theory has the maximal number $N=n$ of degrees of freedom and its constraint structure reads
\begin{equation*}
\fbox{$%
\begin{array}{lll}
\mathrm{First\;class:}\qquad & \bar{G}_{a}=\pi _{a}^{0}\,, & n\,, \\
& G_{a}=D_{i}\pi _{a}^{i}\,,\qquad & n\,.
\end{array}
$}
\end{equation*}
It is interesting that here we have the same set of the first
class constraints as in non-Abelian Yang-Mills theory. The essential
difference between two theories is that the Yang-Mills' Hamiltonian
density contains the usual kinetic term $\frac{1}{2}\pi^2$ which is not a constraint, because the Yang-Mills theory does not have reparametrization symmetry.

With these constraints, the Hamiltonian density (\ref{H}) becomes
\begin{equation}
H=u^{a}\pi _{a}^{0}+v^{a}D_{i}\pi _{a}^{i}\,.
\end{equation}
A time evolution of the variable $F(A,\pi )$ on the phase space is
given by the Poisson bracket
\begin{equation}
\dot{F}=\int d^{2}x'\,\{F,H'\}\,.
\end{equation}
Evolution of the constraints $G_{a}$ and $\bar{G}_{a}$ does not generate  the new
ones,
\begin{eqnarray}
\overset{\cdot }{\bar{G}}_{a} &=&0\,,  \notag \\
\dot{G}_{a} &=&f_{ab\,}^{\ \ c}v^{b}G_{c}=0\,,
\end{eqnarray}
so we have the complete set of them.

Now we have a good Hamiltonian that fulfills all requirements (\textit{i})-(\textit{iv}) and we can
 obtain the Hamiltonian equations,
\begin{equation}
\begin{array}{llll}
\dot{A}_{0}^{a} & =u^{a}\,, & \dot{\pi}_{a}^{0} & =0\,, \\
\dot{A}_{i}^{a} & =-D_{i}v^{a}\,,\qquad & \dot{\pi}_{a}^{i} &
=f_{ab}^{c}v^{b}\pi _{c}^{i}\,.%
\end{array}
\end{equation}
Let us notice that $F_{\mu \nu }^{a}=0$ is not an equation of motion of this
theory for arbitrary multipliers.
For example, $F^a_{0i}=-D_i(v^a+A_0^a)$ and $F^a_{ij}\neq 0$.
According to (\ref{diff-gauge identity}), diffeomorphisms
are not on-shell equivalent to gauge transformations, and the theory is not invariant under general coordinate transformations.

Before moving to the next case, let us analyze this theory and try to get some insight about possible improvements.

From the definition (\ref{action}), the canonical action becomes
\begin{eqnarray}
I[A,\pi ,u] &=&\int dt\,d^{2}x\,\left[ \dot{A}_{\mu }^{a}\pi _{a}^{\mu
}-\left( u^{a}\pi _{a}^{0}+v^{a}D_{i}\pi _{a}^{i}\right) \right]  \notag \\
&=&\int dt\,d^{2}x\,\left[ \left( \dot{A}_{0}^{a}-u^{a}\right) \pi
_{a}^{0}+\left( \dot{A}_{i}^{a}+D_{i}v^{a}\right) \pi _{a}^{i}\right] \,.
\label{actionNoNoNo}
\end{eqnarray}
Now it is explicit that this simple constraints choice does not lead to a
gravitational theory. The action (\ref{actionNoNoNo}) is gauge invariant but
not diffeomorphisms invariant. Vanishing Hamiltonian is, therefore,
necessary, but not sufficient condition for a theory to be diffeomorphism
invariant.

This theory cannot be put in the covariant form without imposing
additional conditions. We can try to determine Hamiltonian multipliers using the
fact that some of the constraints can be primary, and another secondary.
For example, inspired by Yang-Mills theory, we can suppose that
$\bar{G}_{a}=\pi_{a}^{0}$ is a primary constraint and allow the multiplier to depend on the
phase space variables. Then its consistency condition (time evolution)
leads, the same as in  the Yang-Mills case, to the secondary constraint
$G_{a}=D_{i}\pi_{a}^{i}$.  This assumption now gives
\begin{equation}
0=\overset{\cdot }{\bar{G}}_{a}=\int d^{2}x'\,\left( \frac{\partial
u^{\prime b}}{\partial A_{0}^{a}}\,\bar{G}_{b}^{\prime }+\frac{\partial
v^{\prime b}}{\partial A_{0}^{a}}\,G_{b}^{\prime }\right) \,,
\end{equation}%
where the first term vanishes on-shell, as we already know that $\bar{G}_{a}=0$.
The second term yields a secondary constraint $G_{a}$ only if
$\frac{\partial v^{\prime b}}{\partial A_{0}^{a}}\neq 0$. The simplest choice is, again
as in Yang-Mills theory,
\begin{equation}
v^{a}=-A_{0}^{a}\,.
\end{equation}
The sign minus is added for convenience.  This choice enables to have on-shell $F^a_{0i}=0$, but it is still $F^a_{ij}\neq 0$.

The canonical action becomes \cite{Miskovic-Vukasinac}
\begin{eqnarray}
I &=&\int d^{3}x\,\left[ \dot{A}_{i}^{a}\pi _{a}^{i}+A_{0}^{a}D_{i}\pi
_{a}^{i}+\left( \dot{A}_{0}^{a}-u^{a}\right) \pi _{a}^{0}\right]   \notag \\
&=&\int d^{3}x\,\left( F_{0i}^{a}\pi _{a}^{i}-\bar{u}^{a}\pi _{a}^{0}\right)
\,,  \label{actionNoNo}
\end{eqnarray}%
where, in the second line, we redefined  the arbitrary multiplier as $\bar{u}%
^{a}=u^{a}-\dot{A}_{0}^{a}$.

From this form of the action it is easy to identify that non-invariance is due to the missing components $F^a_{ij}$.
Since here $A^a_i$ are dynamical fields carrying degrees of freedom, $F^a_{ij}$ is not a pure gauge and it is not a constraint, so it cannot appear in the action.
Another problem of this action is that the momenta $\pi _{a}^{i}$ cannot be
integrated out to make the action a functional of the gauge field only. In
the case of $\pi _{a}^{0}$ this is not problematic as this field is clearly
unphysical, but dynamical $\pi _{a}^{i}$ needs another treatment.

These two problems of $I$ suggest that a way to improve the theory would be to add the spatial diffeomorphisms
constraints, $K_i$, in the Hamiltonian action, and also
introduce second class constraints that would enforce a relation  $\pi
_{a}^{i}=\mathcal{L}_{a}^{i}(A)$, to help integrate out the momenta in a
covariant way.

We shall explore both possibilities. Let us start from the first option
which introduces spatial diffeomorphisms without involving second class
constraints.

\subsection{Theory containing only first class constraints}

In this section  we assume that spatial diffeomorphisms are
independent first class constraints ($\varepsilon =1$) and there are no
second class constraints, $m=0$. Based on our experience with the gauge
constraints gained in Section \ref{NoNo},  some arbitrary multipliers can be
identified with unphysical gauge field components if we assume that $\bar{G}%
_{a}$ is primary and $G_{a}$ secondary constraint. In this settings, there
are $2n+2$ first class constraints,
\begin{equation*}
\fbox{$%
\begin{array}{lll}
\text{\textrm{Primary first\ class:}}\qquad  & \bar{G}_{a}=\pi _{a}^{0}\,, & n\,, \\
& K_{i}=F_{ik}^{b}\pi_{b}^{k}\,,\qquad  & 2\,, \\
\text{\textrm{Secondary first\ class:}} & G_{a}=D_{i}\pi _{a}^{i}\,. & n\,,
\end{array}
$}
\end{equation*}
satisfying the Poisson brackets (\ref{GG}) and (\ref{GK}). Note that $K_i$  have to be primary
constraints.
The number of physical degrees of freedom is $N=n-2$.

The Hamiltonian density is of the form
\begin{equation}
H=u^{a}\pi _{a}^{0}+v^{a}D_{i}\pi _{a}^{i}+\zeta ^{i}F_{ij}^b\pi_b^j\,,
\end{equation}
where we have to replace $v^{a}=-A_{0}^{a}$ to ensure that the evolution of $%
\bar{G}_{a}=0$ leads to $G_{a}=0$. Equations of motion  read
\begin{equation}
\begin{array}{llll}
\dot{A}_{0}^{a} & =u^{a}\,, & \dot{\pi}_{a}^{0} & =0\,, \\
\dot{A}_{i}^{a} & =D_{i}A_{0}^{a}+\zeta^j F_{ji}^a\,,\qquad  &
\dot{\pi}_{a}^{i} & =f_{ab}^{\ \ c}A_{0}^{b}\pi _{c}^{i}+
D_j(\zeta^j\pi^i_a)-D_j\zeta^i\pi^j_a\,.
\end{array}
\end{equation}

The canonical action becomes
\begin{equation}
I=\int d^{3}x\left( \dot{A}_{\mu }^{a}\pi _{a}^{\mu }-u^{a}\pi
_{a}^{0}+A_{0}^{a}D_{i}\pi _{a}^{i}-\zeta ^{i}F_{ij}^{b}\pi _{b}^{j}\right)
\end{equation}%
or, equivalently, after introducing the multiplier $\bar{u}^{a}=u^{a}-\dot{A}%
_{0}^{a}$, the result is
\begin{equation}
I=\int d^{3}x\left( F_{0i}^{a}\pi _{a}^{i}-\zeta ^{i}F_{ij}^{b}\pi _{b}^{j}-%
\bar{u}^{a}\pi _{a}^{0}\,\right) .
\end{equation}

This theory illustrates a nontrivial gauge theory also invariant under
spatial diffeomorphisms, which possesses dynamical degrees of freedom in
three dimensions. The time-like diffeomorphisms are  absent.
 This action can describe a theory that is not invariant under coordinate transformations in one direction only, such as diffeomorphism invariant theory on a brane or a non-relativistic model. Work on possible applications is currently in progress.

If we want a fully diff-invariant theory, we need some additional ingredients, and one
possibility is to consider a theory with second class constraints. We do it
in the next example.

\subsection{Theory with zero degrees of freedom \label{Zero}}

Let us analyze another extreme case, where a number of degrees of freedom is
zero, $N=0$. Taking into  account the inequality (\ref{N>0}), we can
have $N=0$ when the number of second class constraints $2m$ depends on the
dimension of non-Abelian group $n$ and the number of independent spatial
diffeomorphisms $2\varepsilon $, as
\begin{equation}
m=n-2\varepsilon \,.
\end{equation}%
There are two possibilities: $\varepsilon =0$ and $\varepsilon =1$.
In the first case, we get $m=n$ and according to the table of allowed
multiplets of the second class constraints, this naturally corresponds to
the constraints of the form $\phi _{a}^{i}$.  Alternatively,
we can also choose two sets of the constraints $\phi _{a}$, or $2n$ scalar
constraints, etc, but, in general in these cases it is difficult for
larger $n$ to construct a sufficient number of independent scalars or
vectors satisfying suitable algebra, whereas for $\phi _{a}^{i}$ we need
only one covariant set of the constraints, which clearly becomes the
simplest choice to explore from now on.

Similarly, when $\varepsilon =1$, we need $m=n-2$, and because there is no one constraint which has $2(n-2)$ components, it implies that we have to take
a set of them, for example, $2(n-2)$ scalar constraints. As mentioned before, we will not discuss here these cases.

To conclude, in this section we analyze the theory with zero degrees of
freedom where the {spatial} diffeomorphisms are not an independent symmetry, $%
\varepsilon =0$, and there are $2n$ second class
constraints with the index structure $M=\binom{i}{a}$:%
\begin{equation*}
\fbox{$%
\begin{array}{lll}
\text{\textrm{Primary first class :}} & \bar{G}_{a}=\pi _{a}^{0}\,, & n\,,
\\
\text{\textrm{Primary {or} secondary first class :}} & G_{a}=D_{i}\pi
_{a}^{i}\,,\qquad & n\,, \\
\text{\textrm{Primary {or} secondary second class :}}\qquad & \phi _{a}^{i}\,,
& 2n\,.%
\end{array}%
$}
\end{equation*}

We already saw in previous sections that it is convenient to divide the
constraints into the primary and secondary ones because, in that way, we can
identify some Hamiltonian multipliers as gauge fields. This separation is
arbitrary, so we define that:\medskip

\qquad -\; There are $p$ primary and $n-p$ secondary constraints $G_{a}$;

\qquad -\; There are $q$ primary and $2n-q$ secondary constraints $\phi
^{i}_{a}$.\medskip

There are no tertiary constraints because we look at theories whose Lagrangian equations of motion are at most second order in time derivatives. Furthermore, since the secondary constraints are
obtained from the primary ones by means of the consistency conditions, there
is always equal or fewer number of them,
\begin{equation}
p+q\geq n\,.
\end{equation}%
For covariance, all $G_{a}$ have to be either primary or secondary, and
similarly for $\phi _{a}^{i}$. Thus, there are four possibilities of the pairs $(p,q)$, that are
$(0,0)$, $(n,0)$, $(0,2n)$ and $(n,2n)$. But $p=q=0$ is not allowed because it does not fulfill the above
inequality, and $p=n$ is not allowed because, in this case, the theory does
not have the second generation of first class constraints and
we saw earlier that they help to obtain a covariant transformation law of canonical
variables and identify some arbitrary multipliers with unphysical gauge field components. There remains only one possibility, that is:
\medskip

\qquad \;\; $G_{a}$  are secondary ($p=0$) and $\phi _{a}^{i}$ are primary ($q=2n$) constraints.\medskip

Because we already have canonical representations of the first class constraints, now we focus on
second class ones, $\phi _{a}^{i}$. The primary constraints are always linear in
momenta, and the index structure yields
\begin{equation}
\phi _{a}^{i}=\pi _{a}^{i}+\mathcal{L}_{a}^{i}(A)\,.
\end{equation}%
The constraint $\phi _{a}^{i}$ has to commute on-shell with all the
generators. In particular, for $\bar{G}_{a}$ that gives
\begin{equation}
\{\phi _{a}^{i},\bar{G}_{b}^{\prime }\}=-\frac{\partial \mathcal{L}_{a}^{i}}{%
\partial A_{0}^{\prime b}}\,,
\end{equation}%
which has to vanish on-shell, finding that $\mathcal{L}_{a}^{i}$ does not
depend on-shell on $A_{0}^{a}$,
\begin{equation*}
\mathcal{L}_{a}^{i}=\mathcal{L}_{a}^{i}(A_{j}^{b})\,.
\end{equation*}%
In principle, in the above expression we can also add the term $A_{0}^{b}\sigma
_{ab}^{i}(A_{j}^{c})$, where $\sigma _{ab}^{i}(A_{j}^{c})$ vanishes on-shell
but, as $A_{0}^{i}$ is a non-physical variable (conjugated to $\pi _{i}^{0}=0
$), this would only lead to a redefinition of the corresponding Hamiltonian
multiplier. We shall therefore set $\sigma _{ab}^{i}=0$ without loss of
generality.

The second class constraints also have to satisfy the algebra (\ref{IIclassA}%
), that is%
\begin{equation}
\{\phi _{a}^{i},\phi _{b}^{\prime j}\}=-\frac{\partial \mathcal{L}%
_{b}^{\prime j}}{\partial A_{i}^{a}}+\frac{\partial \mathcal{L}_{a}^{i}}{%
\partial A_{j}^{\prime b}}\equiv \Omega _{ab}^{ij}(x,x^{\prime })\,,
\end{equation}%
where the symplectic matrix $\Omega _{ab}^{ij}$ is invertible on-shell.
$\Omega $ must be antisymmetric under the simultaneous exchange of the
indices $(i,a,x)\leftrightarrow (j,b,x^{\prime })$. In a local theory, $\Omega
_{ab}^{ij}(x,x^{\prime })=\Omega _{ab}^{ij}(x)\delta $, and then the
antisymmetric indices $[ij]$ can be realized through the constant Levi-Civit%
\'{a} tensor $\varepsilon ^{ij}$. The group indices $(ab)$ have to be
symmetric, so they are proportional to the invertible Cartan metric $g_{ab}$%
. It means that an invertible tensor of an appropriate rank reads
\begin{equation}
\Omega _{ab}^{ij}(x,x^{\prime })=k\,\varepsilon ^{ij}g_{ab}\,\delta \,,
\end{equation}%
where $k$ is a non-vanishing real function of the gauge fields $A_{i}^{a}$.
We shall take $k=const\neq 0$ to ensure invertibility in all points of the spacetime manifold. Then we find a particular solution
\begin{equation}
k\,\epsilon ^{ij}g_{ab}\,\delta =\frac{\partial \mathcal{L}_{a}^{i}}{%
\partial A_{j}^{\prime b}}-\frac{\partial \mathcal{L}_{b}^{\prime j}}{%
\partial A_{i}^{a}}\quad \Rightarrow \quad \mathcal{L}_{a}^{i}=-\frac{k}{2}%
\,\varepsilon ^{ij}A_{aj}\,.
\end{equation}

However, it turns out that this choice of second class constraints is not
consistent with the first class character of $G_{a}$, because their brackets
do not vanish,
\begin{equation}
\{\phi _{a}^{i},G_{b}^{\prime }\}=f_{ab}^{c}\phi _{c}^{i}\delta +\frac{k}{2}%
\,\varepsilon ^{ij}g_{ab}\partial _{j}\delta \,.
\end{equation}%
In order to recover the vanishing brackets between $\phi $ and $G$, we
redefine these constraints in the following way. For the gauge constraint $%
G_{a}$, we can take the form found previously in Eq.(\ref{G,barG}), with the
function $h_{a}$ given by (\ref{F,barF}), and $\phi _{a}^{i}$ can be
modified by the addition of the function $s_{a}^{i}\neq 0$, that is,%
\begin{eqnarray}
G_{a} &=&D_{i}\pi _{a}^{i}+h_{a}(A)\,,  \notag \\
\phi _{a}^{i} &=&\pi _{a}^{i}-\frac{k}{2}\,\varepsilon
^{ij}A_{aj}+s_{a}^{i}(A)\,,  \label{G,Phi-new}
\end{eqnarray}%
where $h_{a}$ and $s_{a}^{i}$ do not depend on $A_{0}^{a}$ in order to
commute with $\bar{G}_{a}=\pi _{a}^{0}$. Now we require that the Poisson
brackets of the constraints (\ref{G,Phi-new}) vanish on-shell. As the
result, we find
\begin{equation}
h_{a}=\alpha\,\varepsilon ^{ij}\partial_i A_{aj}\,,\qquad s_{a}^{i}=\frac{%
k-2\alpha }{2}\,\varepsilon ^{ij}A_{aj}\,,  \label{s_a}
\end{equation}%
where $\alpha $ is a real constant. This result matches with Eq.(\ref{h=dA}).
Note that the constant $k$ cancels out in $\phi^i_a$ and the constraints depend only on $\alpha $, namely
\begin{eqnarray}
G_{a} &=&D_{i}\pi _{a}^{i}+\alpha\,\varepsilon ^{ij}\partial_i A_{aj}\,,  \notag \\
\phi _{a}^{i} &=&\pi _{a}^{i}-\alpha\,\varepsilon
^{ij}A_{aj}\,.  \label{G,Phi-new1}
\end{eqnarray}
The algebra of constraints becomes
\begin{equation}
\begin{array}[b]{llll}
\{G_{a},G_{b}^{\prime }\} & =f_{ab}^{\ c}\,G_{c}\,\delta \,, & \{\phi
_{a}^{i},G_{b}^{\prime }\} & =f_{ab}^{\ c}\phi _{c}^{i}\,\delta \,,\medskip
\\
\{\phi _{a}^{i},\phi _{b}^{\prime j}\} & =2\alpha \,\varepsilon
^{ij}g_{ab}\,\delta \,,\qquad  &  &
\end{array}%
\end{equation}
which is the same as before, up to the replacement $2\alpha \leftrightarrow k$.

As shown in Eq.(\ref{H2}), the extended Hamiltonian density does not contain the second class constraints,
and its explicit form is
\begin{eqnarray}
H &=&u^{a}\bar{G}_{a}+ v^{a}G_{a}\,  \notag \\
&=&u^{a}\pi _{a}^{0} + v^{a}\left( D_{i}\pi _{a}^{i}+\alpha \,\varepsilon
^{ij}\partial _{i}A_{aj}\right) \,.
\end{eqnarray}
Similarly as in previous examples, the function $G_{a}$ arises as a secondary constraint if $v^{a}$ is chosen as
\begin{equation}
v^{a}=-A_{0}^{a}\,.  \label{v}
\end{equation}
It implies that the Hamiltonian density read
\begin{equation}
H=u^{a}\bar{G}_{a}-A_{0}^{a}\,\left( D_{i}\pi _{a}^{i}+\alpha \,\varepsilon
^{ij}\partial _{i}A_{aj}\right) \,.
\end{equation}%
Hamiltonian equations of motion are
\begin{equation}
\begin{array}{llll}
\dot{A}_{0}^{a} & =u^{a}\,, & \dot{\pi}_{a}^{0} & =D_{i}\pi _{a}^{i}+\alpha
\,\varepsilon ^{ij}\partial _{i}A_{aj}=0\,, \\
\dot{A}_{i}^{a} & =D_{i}A_{0}^{a}\,,\qquad  & \dot{\pi}_{a}^{i} &
=-f_{ab}^{c}A_{0}^{b}\pi _{c}^{i}+\alpha \,\varepsilon ^{ij}\partial
_{j}A_{0a}\,,%
\end{array}
\label{u}
\end{equation}%
Also, we find that $F_{\mu \nu
}^{a}=0$ on-shell since
\begin{eqnarray}
F_{0i}^{a} &=&\dot{A}_{i}^{a}-D_{i}A_{0}^{a}=0\,,  \notag \\
F_{ij}^{a} &=&\frac{1}{2}\,\varepsilon _{ij}\varepsilon ^{kl}F_{kl}^{a}=%
\frac{1}{2\alpha }\,\varepsilon _{ij}(G^{a}-D_{k}\phi ^{ka})=0\,,
\end{eqnarray}%
in agreement with the fact that the theory does not possess locally
propagating degrees of freedom because the basic field is pure gauge. Furthermore,
the spatial diffeomorphisms become dependent on the gauge transformations,
as discussed before.

Using the expression for the multipliers $v^{a}$ and the
constraints $\phi _{a}^{i}$ to eliminate the non-physical momenta $\pi
_{a}^{i}$ in the canonical action (\ref{action}), we find
\begin{equation}
I=\int d^{3}x\,\left[ \alpha \varepsilon ^{ij}\left( \dot{A}%
_{i}^{a}A_{aj}+A_{0}^{a}F_{ij\,a}\right) -\bar{u}^{a}\pi _{a}^{0}\right] \,,
\end{equation}
where we redefined $\bar{u}^a=u^a - \dot{A}^a_0$.
The first two terms in the previous equation form the action of the Chern-Simons theory,
\begin{equation}
I_{\mathrm{CS}}[A]=\alpha \int d^{3}x\,\varepsilon ^{\mu \nu \rho }\left(
A_{\mu }^{a}\partial _{\nu }A_{a\rho }+\frac{1}{3}\,f_{abc}\,A_{\mu
}^{a}A_{\nu }^{b}A_{\rho }^{c}\right) \,,
\end{equation}%
with $\varepsilon ^{ij}\equiv \varepsilon ^{0ij}$, so that
the canonical action can be put in the form
\begin{equation}
I=I_{\mathrm{CS}}[A]-\int d^{3}x\,\bar{u}^{a}\pi _{a}^{0}\,.
\end{equation}%
The actions $I$ and $I_{\mathrm{CS}}$ are physically equivalent, since their difference is decoupled from the Chern-Simons term and on-shell it is satisfied $\pi _{a}^{0}=0$ and $\bar{u}^a=0$.

It is interesting that, in this procedure, we obtain
$G_{a}$ as secondary constraints, whereas the Hamiltonian analysis of
Chern-Simons action gives different secondary
constraints and $G_{a}$ appear as a linear combination of primary and
secondary ones (see Chapter 6.4 in Ref.\cite{Blagojevic:2002du}).
In our approach, however, we start from the extended
Hamiltonian that includes \textit{all} constraints, both primary and
secondary, and reconstruct the Chern-Simons action in a different, but
physically equivalent way, finding that it is the \textit{unique} action
which can be obtained under given assumptions.

\section{Discussion and outlook} \label{Dis}

In this work we showed that a method based on the Hamiltonian formalism can be successfully applied for construction of theories invariant under general coordinate transformations and possessing a non-Abelian gauge symmetry. Our motivation was to build a gravitational theory coupled to non-Abelian matter or high-spin fields in three dimensions. The procedure was previously applied in two dimensions to obtain Wess-Zumino-Novikov-Witten model \cite{Sazdovic} and its supersymmetric extension \cite{ Miskovic-Sazdovic}.

The key step in the construction was finding a canonical representation of symmetry generators in the phase space, i.e., spatial diffeomorphisms and gauge generators.  The Hamiltonian constraint that generates evolution along a time-like direction and forms the Dirac algebra with the generators of spatial diffeomorphisms, was not included in our approach.  As shown in Ref.~\cite{Hojman}, in order to obtain its canonical representation, some additional assumptions were necessary. In consequence, the models obtained by our method, which are not invariant under time reparametrization, can be used to describe the dynamics of non-relativistic theories or, in case of Hamiltonian evolution along spatial direction, the dynamics of geometries containing membranes (which are diff-invariant submanifolds). Work on possible applications is currently in progress.

In particular, we applied the method in three different cases.
The first model has maximal number of degrees of freedom and minimal number of constraints, but it is a gauge theory that is not diffeomorphism invariant.
In the next example, we considered a model with first class constraints only, and obtained a nontrivial gauge theory also invariant under spatial diffeomorphisms. This theory possesses $n-2$ dynamical degrees of freedom, it is not equivalent to Yang-Mills theory and should be better explored in future. Finally, we studied a system without degrees of freedom, which turned out to be the Chern-Simons theory.

These three examples show that the method, based on symmetries and  the canonical representations of their generators, could be used as a powerful tool for exploring possible dynamics of gauge theories.

The project has many open questions to be addressed in future. The most important one is the inclusion of time-like diffeomorphisms in this formalism, which would guarantee that the final theory contains gravitational field.
Another of the future tasks is to analyze the uniqueness of the canonical representations, i.e., discuss non-equivalent representations of the symmetry generators and the resulting canonical actions. For example, one possible extension would be to include primary constraints non-linear in momenta, present in case of free non-relativistic point particle.

An important challenge whose solving is currently in progress, is construction of a physically interesting example of a theory that possess physical degrees of freedom in three dimensions, but is also invariant under spacetime diffeomorphisms. An additional extension concerns an issue of boundary terms, essential for having a well-defined theory in spacetime regions with boundary and definition of conserved quantities, and should also be addressed later on.

\section*{Acknowledgments}

The authors would like to thank M\' aximo Ba\~{n}ados, Alejandro Corichi, Marija Dimitrijevi\'c, Hern\'{a}n Gonz\'{a}lez,
Sergio Hojman, Milica Milovanovi\'c and Jorge Zanelli for useful comments and discussions. This
work was supported in parts by Mexican CIC, UMSNH project and Chilean FONDECYT projects N$^{\circ}$1170765 and N$^{\circ}$1190533.

\end{document}